# Enabling Aloha-NOMA for Massive M2M Communication in IoT Networks

Mohamed Elkourdi, Asim Mazin, Eren Balevi, and Richard D. Gitlin, *Life Fellow, IEEE*
Innovation in Wireless Information Networking Lab (*i*WINLAB)
Department of Electrical Engineering, University of South Florida
Tampa, Florida 33620, USA
E-mail: {elkourdi, asimmazin, erenbalevi}@mail.usf.edu, richgitlin@usf.edu

*Abstract*— The Internet of things (IoT), which is the network of physical devices embedded with sensors, actuators, and connectivity, is being accelerated into the mainstream by the emergence of 5G wireless networking. This paper presents an uncoordinated non-orthogonal random access protocol, an enhancement to the recently introduced Aloha-NOMA protocol, which provides high throughput, while being matched to the low complexity requirements and the sporadic traffic pattern of IoT devices. Under ideal conditions it has been shown that Aloha-NOMA, using power-domain orthogonality, can significantly increase the throughput using SIC (Successive Interference Cancellation) to enable correct reception of multiple simultaneous transmitted signals. For this ideal performance, the enhanced Aloha-NOMA receiver adaptively learns the number of active devices (which is not known *a priori*) using a form of multi-hypothesis testing. For small numbers of simultaneous transmissions, it is shown that there can be substantial throughput gain of 6.9 dB relative to pure Aloha for 0.25 probability of transmission and up to 3 active transmitters.

## I. Introduction

The rapid growth of both the number of connected devices and the data volume that is expected to be associated with the IoT applications, has increased the popularity of Machine-to-Machine (M2M) type communication within 5G wireless communication systems [1]. Uncoordinated random access schemes have attracted lots of attention in the standards of cellular network as a possible method for making massive number of M2M communication possible with a low signaling overhead [2], [3]. This paper advances a novel MAC protocol, dubbed Aloha-NOMA, which is matched to IoT/M2M applications and which is scalable, energy efficient and has high throughput.

The protocol exploits the simplicity of Aloha and the superior throughput of non-orthogonal multiple access (NOMA) [4] and its ability to resolve collisions via use of successive interference cancellation (SIC) receiver [5],[6]. The recently introduced Aloha-NOMA protocol [7] and subsequent enhancements [8] are a promising candidate MAC protocol that can be utilized for low complexity IoT devices. In [8] NOMA is applied to multichannel slotted Aloha to enhance the throughput with respect to conventional multichannel slotted Aloha [9] without the need for any bandwidth expansion. However, a strict slot synchronization is needed in [8] which might make it not suitable for heterogeneous IoT networks.

The Aloha-NOMA protocol is a promising method for not requiring any scheduling, apart from frame synchronization, in which all IoT devices transmit to the gateway at the same time on the same frequency band but also it is energy efficient and has high throughput as will be demonstrated in this paper. This paper presents an enhancement to the Aloha-NOMA protocol where the receiver adaptively learns the number of active devices (which is not known *a priori*) using a form of multi-hypothesis testing [10].

The contributions of this paper can be summarized as follows. First, an enhancement to the Aloha-NOMA protocol is proposed to make it more suitable for IoT applications. Following that, a form of multiple hypotheses testing is exploited to detect the number of active IoT devices and adjust the SIC power levels. Simulation results are performed to show that the enhanced Aloha-NOMA protocol significantly outperforms conventional Aloha in terms of throughput.

The paper is organized as follows. Section II discusses the proposed Aloha-NOMA protocol, Section III demonstrates the use of multiple hypotheses testing for detecting the number of active IoT devices and Section IV proposes a dynamic frame structure for Aloha-NOMA. In Section V, simulation results are presented to show the superiority of the proposed method regarding throughput with respect to the pure Aloha. The paper is concluded with some remarks in Section VI.

## II. Aloha-Noma Protocol For IoT Applications

NOMA has emerged as a promising technology in 5G networks for many applications [4], and the Aloha-NOMA protocol is a synergistic combination of the low complexity Aloha protocol with the high throughput feature of NOMA. The main bottleneck of Aloha systems is the low throughput caused by the high number of collisions, which can be addressed by NOMA. In Aloha-NOMA the signaling overhead is reduced in the detection phase of the proposed protocol where the number of active IoT devices are detected by the gateway using a form of multiple hypotheses testing, which is further explained in Section IV. It is also an energy efficient protocol due to the fact that a SIC receiver resolves collisions, and thus minimizes retransmission. Lastly, the proposed scheme increases the conventional pure Aloha throughput significantly.

The Aloha-NOMA protocol can be suitable for various scenarios where many IoT devices are transmitting simultaneously on the same frequency with different power levels to an IoT gateway and then the received signals can be separated via use of SIC receiver. A sample illustration of this scenario is depicted in Fig. 1 as a smart home with an IoT network. In this model, IoT devices send their data to the IoT gateway whenever they want using the Aloha-NOMA protocol and the IoT gateway distinguish the signals with SIC receiver.

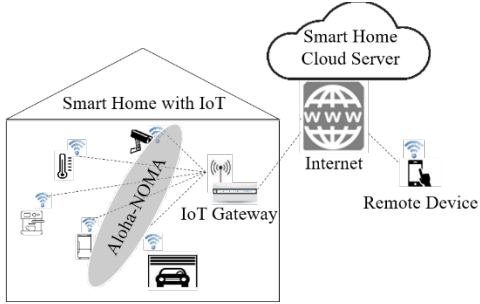

Fig. 1. A use case of Aloha-NOMA in the smart home with IoT.

It is an energy efficient protocol due to the fact that a SIC receiver resolves collisions, and thus minimizes retransmission. Lastly, the proposed scheme increases the conventional pure Aloha throughput significantly.

### III. FLEXIBLE FRAME STRUCTURE FOR THE ALOHA-NOMA PROTOCOL

One of the main practical challenges for enabling the practical implementation of the Aloha-NOMA protocol is the assignment of proper power levels for the IoT devices before transmitting the information; otherwise the signals received from different IoT devices cannot be extracted successively from the composite received signal. This issue becomes more challenging in dynamic environments where the number of IoT devices with information ready to send is continuously changing.

In this section, this challenge is addressed via a flexible frame structure. Such a scheme provides great flexibility in adapting to changing network environments. This structure is opposite to that of TDMA or FDMA in which a new user arrival can completely change the overall frame structure such the additional user must be assigned at least one slot within the frame.

As illustrated in Fig.2, the proposed frame structure is composed of 5 phases. In the first phase, the IoT gateway transmits a beacon signal to announce its readiness to receive packets. Next, the IoT devices with packets ready to transmit send a training sequence to aid the gateway in detecting the number of active IoT devices in the medium. The IoT gateway detects the number of devices requesting transmission via a form of multiple hypotheses testing, as further explained in Section IV, and adjusts the degree[1] of SIC receiver for the optimum power levels. In practice, the SIC receiver has a fixed range of optimum power levels (e.g. $m = 2, 3$). If the IoT devices are registered with the gateway instead of using multi-hypothesis testing, implementation would be simpler, however, this will significantly increase the length of the control phase and thus decrease the payload or throughput considering the potentially large number of IoT devices. In the third phase, if the detected number of active IoT devices is not in the range of the optimal power levels the IoT gateway aborts the transmission and starts the frame again by sending a new beacon signal and implying that the active transmitters use a random backoff. If the detected number of devices is in range, the IoT gateway broadcast the degree of SIC to the transmitters and then each active IoT device randomly picks one of the optimum power levels. If the choices are distinct the SIC receiver can decode the self-identifying signals (device ID + payload) and then the gateway sends an ACK. However, if the active IoT devices did not select distinct power levels, the reselection process is repeated and after a few attempts if there is no successful transmission, the users receive a NACK and enter a random back-off mode. It should be clear, that the proposed protocol will be most efficient when there is a small number of active devices, so that the probability of using the optimum power levels remains high.

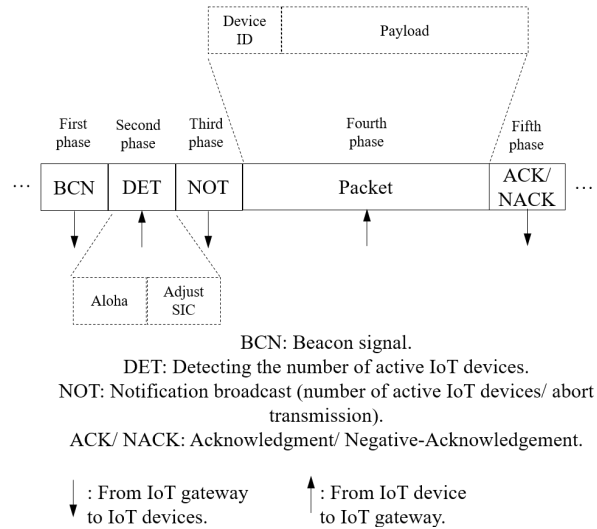

Fig. 2. The proposed protocol frame structure.

### IV. MULTIPLE HYPOTHESIS TESTING: DETECTING THE NUMBER OF ACTIVE IOT DEVICES

This section presents a form of multiple hypothesis testing to detect the number of active devices. The detection of active devices starts in the second phase of the Aloha-NOMA frame as presented in Fig.2.

After receiving the beacon, all the IoT devices send at the same power level a training sequence of length $L$ using the Aloha protocol. The superposed received signal at the IoT gateway from $N$ active transmitting IoT devices is given by

---

[1] We denote a SIC receiver that can process $m$ signals as SIC($m$) and we refer to $m$ as the SIC degree and $m$ is adjusted to agree with the result of the multi-hypothesis test described above.

$$y = Hs + w, \quad (1)$$

where $H = [h_1, h_2, ..., h_N] \in \mathbb{R}^{1 \times N}$ and $h_n$ is the channel gain between the $n$th IoT device and the IoT gateway. $s \in \mathbb{R}^{N \times L}$ is the transmit sequence (e.g. BPSK) from $N$ IoT active devices and $w \in \mathbb{R}^{1 \times L}$ is the additive white Gaussian noise with zero mean and variance $\sigma^2$. The multiple hypothesis test is used to detect the number of $N$ active IoT devices from the total $M$ IoT devices. The following procedure is used to sequentially detect the number of active devices

$\mathcal{H}_0$: Received signal contains only noise
$$y = w$$
$\mathcal{H}_1$: Received signal contains data from at least one IoT device
$$y = h_1 s_1 + w, \quad (2)$$
$\mathcal{H}_N$: Received signal contains data from at most $N$ IoT devices
$$y = Hs + w$$

We assume $h_n = 1, \forall n \in \{1, 2, ..., N\}$. Following the Neyman-Pearson (NP) test, we can write the Likelihood Ratio (LR) testing [10] $\mathcal{H}_N$ vs. $\mathcal{H}_{N-1}$ as

$$\frac{p(y; \sum_{n=1}^{N} s_n, \mathcal{H}_N)}{p(y; \sum_{n=1}^{N-1} s_n; \mathcal{H}_{N-1})} = \frac{exp\left[-\frac{1}{2\sigma^2}(y - \sum_{n=1}^{N} h_n s_n)^T (y - \sum_{n=1}^{N} h_n s_n)\right]}{exp\left[-\frac{1}{2\sigma^2}(y - \sum_{n=1}^{N-1} h_n s_n)^T (y - \sum_{n=1}^{N-1} h_n s_n)\right]} \gtrless_{\mathcal{H}_{N-1}}^{\mathcal{H}_N} \gamma, N = 1, .... M \quad (3)$$

where $s_n \in \mathbb{R}^{1 \times L}$ is the transmitted sequence from the $n^{th}$ IoT device. By taking the logarithm, (3) is simplified to

$$T(y) = \frac{1}{L} \sum_{l=0}^{L-1} y \gtrless_{\mathcal{H}_N}^{\mathcal{H}_{N-1}} \frac{2\sigma^2 \ln \gamma - ((\sum_{n=1}^{N-1} h_n s_n)(\sum_{n=1}^{N-1} h_n s_n)^T + (\sum_{n=1}^{N} h_n s_n)(\sum_{n=1}^{N} h_n s_n)^T)}{-2(\sum_{n=1}^{N-1} h_n s_n + \sum_{n=1}^{N} h_n s_n)} = \gamma' \quad (4)$$

The NP detector, or the test statistic, in (4) compares the sample mean of the received signal to the threshold $\gamma'$ to decide on a hypothesis $\mathcal{H}_N$ or $\mathcal{H}_{N-1}$. The NP test terminates if the number of detecting devices exceeds the SIC receiver optimum power levels, which are 3 levels in this paper. To compute the threshold $\gamma'$ in (4) for a desired probability of false alarm $P_{FA}$, which occurs when deciding $\mathcal{H}_N$ if the test in (4) is greater than the threshold $\gamma'$, so that $P_{FA}$ can be written as

$$P_{FA} = P(T(y) > \gamma'; \mathcal{H}_N) \quad (5)$$

Since the test in (4) under both hypothesis is a Gaussian distribution, that $T(y) \sim \mathcal{N}(\sum_{l=0}^{L-1} \sum_{n=1}^{N-1} s_n, \frac{\sigma^2}{L})$ under $\mathcal{H}_{N-1}$ and $T(y) \sim \mathcal{N}(\sum_{l=0}^{L-1} \sum_{n=1}^{N} s_n, \frac{\sigma^2}{L})$ under $\mathcal{H}_N$ we rewrite (5) as

$$P_{FA} = Q\left(\frac{\gamma' - \sum_{l=0}^{L-1} \sum_{n=1}^{N} s_n}{\sqrt{\sigma^2/L}}\right) \quad (6)$$

Thus, the threshold $\gamma'$ is given by

$$\gamma' = Q^{-1}(P_{FA})\sqrt{\sigma^2/L} + \sum_{l=0}^{L-1} \sum_{n=1}^{N} s_n. \quad (7)$$

Following the same steps, the probability of detecting the number of active devices is

$$P_D = Q\left(\frac{\gamma' - \sum_{l=0}^{L-1} \sum_{n=1}^{N-1} s_n}{\sqrt{\sigma^2/L}}\right) \quad (8)$$

From (7), (8) we can write the $P_D$ as a function of energy to noise ratio as

$$P_D = Q\left(Q^{-1}(P_{FA}) + \frac{\sum_{l=0}^{L-1} \sum_{n=1}^{N} s_n - \sum_{l=0}^{L-1} \sum_{n=1}^{N-1} s_n}{\sqrt{\sigma^2/L}}\right) \quad (9)$$

The probability of correct detection of the number of active users as a function of the SNR for $P_{FA} = 0.1$ is shown in Fig.3. Observe that the detection performance increases monotonically and smoothly with increasing SNR.

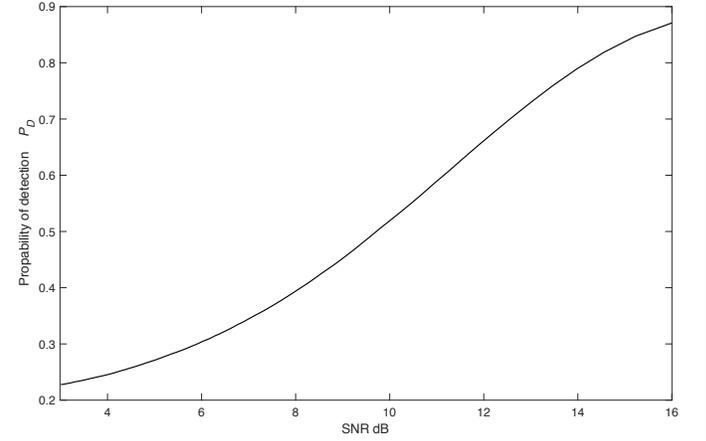

Fig.3 Detection probability of the number of active devices as a function of SNR.

## V. SIMULATION RESULTS

In this section, the simulation results are presented to evaluate the throughput performance of Aloha-NOMA protocol. Throughout the simulation, we assume there is one IoT gateway with a single antenna and a total of $M$ IoT devices. A binomial distribution is considered to model the random number of active IoT devices $N$, each with probability of transmission $p_T$.

$$P_r(N; p_T, M) = \binom{M}{N} p_T^N (1 - p_T)^{M-N} \quad (10)$$

Fig. 4 shows the throughput of Aloha-NOMA for different values of $p_T$, $M = 10$ and $k = 3$ attempts for the random selection of distinct optimum power levels. Throughput is the number successful transmissions for each probability of transmission. As expected, we observe that the throughput decreases with increasing probability of transmission. More importantly, we can see that the throughput of the Aloha-NOMA protocol is always higher than that of the pure Aloha protocol. In particular, when the probability of transmission is 0.25, the throughput of Aloha-NOMA with 3 power levels becomes almost 5 times higher than that of (conventional) pure Aloha. This demonstrates that NOMA with a SIC receiver can help improve the throughput of pure Aloha, especially when the

IoT devices are active.

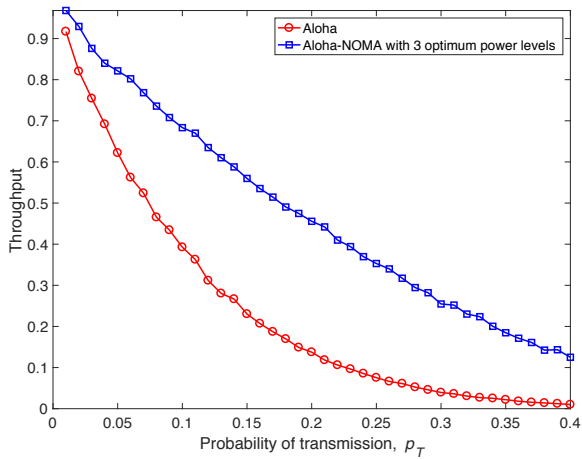

Fig. 4. Throughput of Aloha vs. Aloha-NOMA for different values of probability of transmission and different optimum power levels ($m$=2, 3) when $M$=10 and $k$=3.

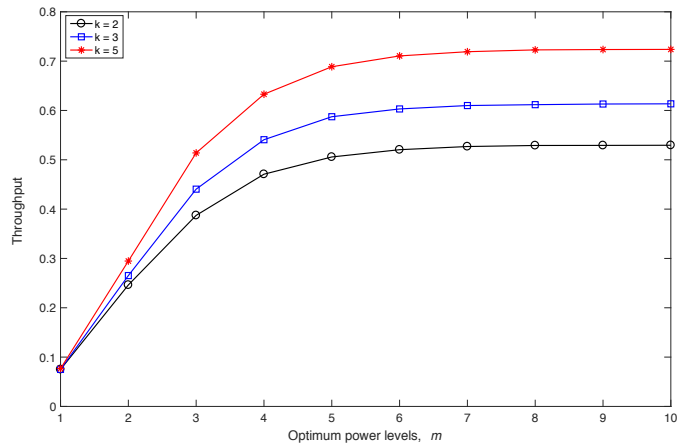

Fig. 5. Throughput of Aloha-NOMA for different optimum power levels when $k$= 2, 3 and 5 and $M$=10.

In order to see the impact of the number of optimum power levels on the throughput of Aloha-NOMA, we show the throughput of Aloha-NOMA for different power levels in Fig. 5. For $M$=10, $k$=3 and probability of transmission of 0.25, the throughput of Aloha-NOMA increases with the increase in optimum power levels (SIC degree). For example, Aloha-NOMA with 3 power levels has a higher throughput than Aloha-NOMA with 2 power levels. However, the throughput improvement becomes insufficient for optimum power levels greater than 5 (saturation in the throughput gain). Also, the simulation results in Fig. 5, shows the impact of the number of attempts $k$, for picking distinct optimum power levels, on throughput for different optimal power levels. The more attempts allowed for picking the optimum power levels, the higher the throughput that can be achieved at the cost of increased delay. However, the delay analysis is beyond the scope of this paper.

## VI. CONCLUDING REMARKS

This paper is directed towards an enhancement of the recently proposed Aloha-NOMA MAC layer protocol that is easy to implement, energy efficient, scalable and compatible with the low complexity requirements of IoT devices. The synergistic combination of Aloha protocol with NOMA and SIC receivers was demonstrated to significantly improve the throughput performance with respect to the Aloha protocol. This paper enhances the Aloha-NOMA protocol by providing the IoT gateway a means to determine the number of active IoT devices in the medium. Knowing the number of active IoT devices is essential in order to optimize the SIC power levels and the ability to distinguish between signals from different IoT devices transmitting on the same time and frequency. In the case of massive number of active IoT devices the power levels become high, whereas the IoT devices usually have a limited power capabilities. Therefore, in this paper the number of active devices that can achieve a successful transmission using SIC is limited to keep the power levels modest. It was shown that with correct detection, there is a substantial throughput gain of 6.9 dB relative to pure Aloha for 0.25 probability of transmission and up to 3 active transmitters.